\documentclass[aps,preprintnumbers,amsmath,amssymb]{revtex4}

\usepackage{graphicx}
\usepackage{xcolor}  
    
\DeclareMathOperator{\Tr}{Tr}

\begin{document}
\title{Langevin equation with potential of mean force: The case of anchored bath}
\author{Alex V. Plyukhin}
\email{aplyukhin@anselm.edu}
 \affiliation{ Department of Mathematics,
Saint Anselm College, Manchester, New Hampshire 03102, USA 
}

\date{\today}

\begin{abstract}
The potential of mean force (PMF) is an effective average potential acting on an open system, renormalized due to the interaction with the surrounding thermal bath. The PMF determines the correction to the equilibrium Gibbs distribution, but it is generally unclear 
how to implement the concept for the time-dependent phenomena
described by a (generalized) Langevin equation.
We study a model where the system is a single particle (so there are no complications related to internal forces) and 
a non-trivial PMF is due to the presence of on-site (anchor) potentials applied to the bath particles. 
We found that the PMF does not merely replace  
the external potential, but also makes the dissipation kernel and statistical properties of noise dependent on the system's position.
That dependence is determined by the internal bath and system-bath interactions and is a priori unknown.  Therefore, in the general case 
the Langevin equation with the PMF is not closed and thus inoperable. 
However, for systems with linear forces the aforementioned dependence on the system's position is canceled. As an example, we consider a model where the bath is formed by the Klein-Gordon chain, i. e. a harmonic chain  with on-site harmonic potentials. In that case, the generalized 
Langevin equation has the standard form with an external  potential replaced by a quadratic PMF.  

\end{abstract}

\maketitle
\section{Introduction}

When addressing the dynamics of open systems, one often assumes that the environment, or the thermal bath, is passive  
in the sense that it does not produce any {\it systematic} force
on the system kept at rest.
A passive bath is still responsible for  the velocity-dependent dissipative
force on the moving system,  but the only force it exerts on the {\it fixed} system
is Langevin noise that vanishes after an appropriate averaging.
The passivity of a thermal bath does not necessarily imply  that  the system-bath interaction is weak, but often holds due to symmetry. 
The required symmetry is not a property of the bath alone, but of the bath and system combined. For example, in conceptual models of Brownian motion the system is a single point-like particle, the bath is a homogeneous fluid, and a potential describing the system-bath  
interaction is pairwise. For that setting, the average bath-induced force on a fixed system vanishes because contributions from different bath molecules are canceled out.    As a result, the bath is passive for an arbitrary strength of the system-bath interaction. 

Now, in the above example, let us replace the point-like particle by a dumbbell-like system consisting of two particles, say $1$ and $2$, and   consider the force the bath acts on particle $1$. That force generally does not vanish  after averaging because particle $1$ interacts with the bath whose symmetry is broken  by the presence of particle $2$.
Generalizing, for any system consisting of more than one particle, the assumption of the passive bath does not generally hold because one part of the system breaks the symmetry of the bath interacting with other parts of the system.
Under certain conditions, the effect may be negligible, but for many systems studied in chemical physics  
it is often essential~\cite{Kirkwood,DO,Roux}.

In this paper, we consider a model
in which molecules of the bath experience, in addition to pairwise potentials from other molecules,  also individual on-site potentials. That potentials, tending to localize bath molecules, break translational invariance of the bath. As a result, the bath is non-passive
even when the system is  
a single particle.

When the bath is passive, 
the equilibrium distribution for the system 
has the canonical form
\begin{eqnarray}
    \rho_s=\frac{1}{Z_s}\, e^{-\beta H_s}, \quad 
    Z_s=\Tr_s\left(e^{-\beta H_s}\right),
    \label{rho_c}
\end{eqnarray}
where $H_s$ is the bare Hamiltonian of the system and $\Tr_s(\cdots)$ denotes integration over the sets of the system's coordinates $X$ and momenta $P$ (we assume that the system is classical). For a non-passive bath $\rho_s$ involves the re-normalized Hamiltonian of the system $H_s^*$, called the Hamiltonian of mean force (HMF)
~\cite{Kirkwood,DO,Roux,Gelin,Lutz,Talkner}  
\begin{eqnarray}
    \rho_s=\frac{1}{Z_s^*}\, e^{-\beta H_s^*}, \quad 
    Z_s^*=\Tr_s\left(e^{-\beta H_s^*}\right).
    \label{rho_sg}
\end{eqnarray}
In classical models, 
the renormalization affects only the system's potential energy $V(X)$, adding to it a bath-induced correction $\Delta V$. 
Then the HMF can be written as
\begin{eqnarray}
    H_s^*=K(P)+V_*(X),
\end{eqnarray}
where $K(P)$ is the system's kinetic energy and $V_*(X)$ is the renormalized potential
\begin{eqnarray}
    V_*(X)=V(X)+\Delta V(X),
    \label{pmf_def0}
\end{eqnarray}
called the potential of mean force (PMF). 
The bath-induced potential $\Delta V$ 
in general depends on temperature and quantifies the 
deviation of the equilibrium distribution from the canonical form (\ref{rho_c}).

As a note on nomenclature, the term PMF is sometimes reserved to $\Delta V$, or even to the total renormalized Hamiltonian of the system $H_s^*$ (thus  identifying the PMF and the HMF). However, these definitions do not appear to be very common.  Our definition  of the PMF as  a total effective potential acting on the system, Eq. (\ref{pmf_def0}), is the same as in Ref.~\cite{Talkner}. A drawback of such terminology  is that it does not offer a generally-accepted term for the key quantity $\Delta V$. We shall refer to $\Delta V$ as the bath-induced potential and call  the bath non-passive if it generates a non-zero (and non-constant) $\Delta V$. We avoid the term ``active bath" which may  suggest irrelevant connotations with the realm of active matter, which is beyond the scope of the paper.

As discussed in the next section, the PMF is defined as a quantity that characterizes the equilibrium distribution and determines average equilibrium properties of the system.   
Clearly, the PMF must also affect time-dependent properties, such as time correlation and  response functions, described by a Langevin equation - but {\it how} it affects them is not always obvious. The simplest assumption is that the Langevin equation and the corresponding  fluctuation-dissipation relation retain their standard form and the only required modification is the replacement of the bare system potential  by the PMF.
That approach, broadly adopted in empirical studies, received theoretical endorsement in 
Refs.~\cite{Lange,Kinjo}, where the standard generalized Langevin equation with the PMF was derived 
from the underlying dynamics using the projection operator technique. 
However, those  derivations were  later 
criticized in Refs.~\cite{Schilling,Schilling_review}, with the conclusion that for systems with a PMF the standard Langevin equation 
can be derived neither exactly nor 
using any well-controlled approximation.
On the other hand, in Refs.~\cite{Vroy,Vroy2,Lee} it was argued that a Langevin equation actually can be recovered from the first principles, but only at the expense of the noise and the dissipation kernel being position-dependent and with a modified fluctuation-dissipation relation. That conclusion seems to corroborate the earlier results for a system of several Brownian particles~\cite{DO}. Using a different approach, 
a generalized Langevin equation with an 
additional dissipation force, independent of the system's velocity, was suggested in~\cite{Netz}.

In all studies cited above,  the derivation of the Langevin equation with a PMF
was addressed for many-particle systems with internal  forces, which significantly complicates the analysis. 
In this paper we consider a model where the system is a single Brownian particle with no internal degrees of freedom. In that case, as we mentioned above,  a translationally-invariant bath is passive 
due to symmetry, and thus the PMF is  redundant.
In our model, the translational invariance of the bath is broken by on-site (or anchor) potentials applied to the particles of the bath.  Such a bath is not passive; it exerts a systematic average force on a  fixed system. Our goal is to determine how this force affects the Langevin equation for the system.

Our interest in such a model is threefold.
First, the model involves a non-trivial PMF but has no complications related to internal forces in the system. 
Second, the Langevin dynamics with an anchored bath is relevant for a variety of physical systems, see~\cite{Savin,Flach,Prem,Dhar,Flach2} and references therein. For example, in complex molecular structures like proteins, the  lighter  and heavier units may be interpreted as the bath degrees of freedom and anchor (substrate) centers, respectively~\cite{Kneller,Xie,Milster}.
Third, the spectrum of an anchored bath has both upper and lower bounds, which may lead to peculiar ergodic properties  of the Langevin dynamics~\cite{Plyukhin_NEO}.

We shall consider the model first for the general and then for a more specific settings. For the general setting, we make no assumptions about specific forms of the potentials involved.
For that general case, we are unable to derive a generalized Langevin equation in the standard form.  The equation we derived instead involves noise and the dissipation kernel, which depend on the system's position in a way which is a priori unknown. As a result, the equation is not closed and practically unmanageable.

On the other hand, 
the situation is simplified significantly 
for a specific setting when the bath is formed by 
the harmonic Klein-Gordon chain. Due to the linearity of that model, the aforementioned dependence of the dissipation kernel on the system's position is canceled, and 
the generalized Langevin equation has the standard form  with a quadratic PMF.

The layout of the paper is as follows. In Section 2 we review the PMF.
Although the concept is quite mature, there is some ambiguity and variations in the literature, so we feel a brief review might be useful.
In Section 3 we outline the model with
a single-particle system and anchored bath. In Sections 4 and 5 we attempt to derive the Langevin equation for the general setting using the Mazur-Oppenheim projection operator technique. In Section 6  we
describe a more specific setting with the Klein-Gordon lattice as a bath and evaluate the corresponding PMF.  In Section 7 we show that for the Klein-Gordon model the Langevin equation derived  in Section 5 is reduced to the standard form. In Appendix we derived that equation by direct integration of bath variables.

\section{Review of PMF}

Consider a classical  system, with bare Hamiltonian $H_s(X,P)$, interacting with a thermal bath with  bare Hamiltonian $H_b(x,p)$ via potential $\Phi(X,x)$.  Here $(X,P)$ and  $(x,p)$
denote the sets of coordinates and momenta for the system and the bath, respectively.  
The total Hamiltonian  $H=H_s+H_b+\Phi$
is convenient to compose as 
\begin{eqnarray}
    H=H_s+H_0(X),\quad H_0(X)=H_b+\Phi,
    \label{H}
\end{eqnarray}
where $H_0(X)$ is the Hamiltonian of the bath interacting with the system with fixed position vector $X$. 

Suppose the overall system (the system and the bath combined) is in thermal equilibrium with canonical distribution
\begin{eqnarray}
    \rho=\frac{1}{Z}\,e^{-\beta H}=\frac{1}{Z}\,e^{-\beta (H_s+H_0)}, \quad Z=\Tr_{sb}\left( e^{-\beta H}\right),
    \label{rho}
\end{eqnarray}
where $\Tr_{sb}(\cdots)$ denotes integration over phase-space variables of 
both the system and the bath. We are interested in the reduced (marginal) distribution $\rho_s$ for the system only,
\begin{eqnarray}
    \rho_s=\Tr_b(\rho)=
    \frac{Z_0(X)}{Z}\,e^{-\beta H_s}.
\label{aux0}
\end{eqnarray}
Here $\Tr_b(\cdots)$ denotes integration over phase-space variables of 
the bath, and  $Z_0(X)$ is the partition function for the canonical distribution $\rho_0$ corresponding to $H_0$, 
\begin{eqnarray}
   \rho_0=\frac{1}{Z_0(X)}\, e^{-\beta H_0}, \quad 
    Z_0(X)=\Tr_b\left(e^{-\beta H_0}\right).
    \label{rho_0}
\end{eqnarray}
Distribution $\rho_0$  is the  equilibrium distribution for the bath connected to the system fixed at position $X$, so that  
partition function $Z_0$ in general depends on $X$. 
Since that dependence is a priori unknown, expression (\ref{aux0}) for $\rho_s$
is not directly useful or evocative. For some special settings the total partition function can be, exactly or approximately, factorized, $Z=Z_s\,Z_0$.  In that case,
$Z_0$ is canceled out in Eq. (\ref{aux0}), and $\rho_s$  takes the Boltzmann-Gibbs form (\ref{rho_c}),
\begin{eqnarray}
    \rho_s=\frac{1}{Z_s}\, e^{-\beta H_s}, \quad 
    Z_s=\Tr_s\left(e^{-\beta H_s}\right).
    \label{rho_cc}
\end{eqnarray}
However, in the general case the aforementioned factorization of $Z$ does not hold, and expression (\ref{aux0}) for $\rho_s$ is not reduced to
the form (\ref{rho_cc}). Yet it is possible to present the general 
expression (\ref{aux0}) in a more suggestive form as follows.

Integrating over the bath variables, the total partition function can be written as
\begin{eqnarray}
    Z=\Tr_{sb}[e^{-\beta (H_s+H_0)}]=\Tr_s[e^{-\beta H_s}\, Z_0(X)],
    \label{Z}
\end{eqnarray}
then Eq. (\ref{aux0})
takes the form
\begin{eqnarray}
    \rho_s=\frac{Z_0(X)}{\Tr_s[e^{-\beta H_s}\, Z_0(X)]}\, e^{-\beta H_s}.
{\label{rho_s_aux}}
\end{eqnarray}
It is convenient to parameterize the dependence of $Z_0$ on $X$  as 
\begin{eqnarray}
    Z_0(X)=c\,e^{-\beta\Delta V(X)}.
\label{aux2}
\end{eqnarray}
While $Z_0$ has no direct physical meaning, the new function
\begin{eqnarray}
    \Delta V(X)=-\frac{1}{\beta}\,\ln\left[\frac{Z_0(X)}{c}\right],
    \label{DV_def}
\end{eqnarray}
can be 
interpreted as an effective bath-induced  potential acting on the system in equilibrium.
Indeed, substituting (\ref{aux2}) into (\ref{rho_s_aux}) gives
\begin{eqnarray}
    \rho_s=\frac{1}{Z_s^*}\,e^{-\beta\, H_s^*}, \quad Z_s^*=\Tr_s[e^{-\beta H_s^*}]
\label{rho_s}
\end{eqnarray}
where 
\begin{eqnarray}
    H_s^*=H_s+\Delta V(X)
    \label{HMF_def}
\end{eqnarray}
is called the Hamiltonian of mean force (HMF). By writing the bare  Hamiltonian of the system as the sum of  kinetic and potential energies 
$H_s=K(P)+V(X)$,
the HMF can be presented as
\begin{eqnarray}
    H_s^*=K(P)+V_*(X), \quad V_*(X)=V(X)+\Delta V(X),
\label{PMF}
\end{eqnarray}
The renormalized potential $V_*(X)$ is commonly referred to as the potential of mean force (PMF). 
It is the sum of the potential energy  $V$ of the bare system (which describes both the internal  and external forces for the system disconnected from the bath) 
and the bath-induced  potential $\Delta V$.

One may find slightly different definitions of the PMF in the  literature. In particular, some authors identify the PMF with the bath-induced potential $\Delta V$ and do not include in the PMF the potential of the bare system $V$~\cite{DO}.  Our definitions of the HMF and the PMF by Eq. (\ref{PMF}) coincide  with those in Ref.~\cite{Talkner}. There are also  differences in the literature about specific choices for parameter $c$ in Eq. (\ref{aux2});
we shall discuss that below.

The  interpretation of $\Delta V$ as the bath-induced effective potential can be further specified considering the force $F_\alpha$, exerted by the bath on $\alpha$-th particle of the system
with the position vector $X_\alpha$,
\begin{eqnarray}
  F_\alpha=-\frac{\partial H_0}{\partial X_\alpha}=-\frac{\partial \Phi}{\partial X_\alpha}.
\end{eqnarray} 
Consider the average force $\langle F_\alpha\rangle$,
where the average is taken over bath variables with distribution $\rho_0$, Eq. (\ref{rho_0}), 
\begin{eqnarray}
    \langle F_\alpha\rangle=\Tr_b(\rho_0\, F_\alpha)=
    \frac{1}{Z_0(X)}\,\int  dx \,dp \,e^{-\beta H_0}\, F_\alpha.
    \label{F_av_def}
\end{eqnarray} 
Differentiating Eq. (\ref{DV_def}), one finds 
\begin{eqnarray}
\langle F_\alpha\rangle= -\frac{\partial}{\partial X_\alpha}\,\Delta V(X).
\label{F_av}
\end{eqnarray}
Therefore,  $\Delta V$ can be interpreted as a potential of the average force exerted by the bath on the fixed system. 
As was mentioned in the Introduction, for a one-particle system in a fluid 
that force  vanishes  by symmetry, but in the general case it is not zero.


Note that the total average force on the $\alpha$-th particle of the system, i. e. the sum of internal, external, and bath-induced forces,
\begin{eqnarray}
    \langle F_\alpha^{tot}\rangle =-\left\langle \frac{\partial H}{\partial X_\alpha}\right\rangle=-\frac{\partial V}{\partial X_\alpha}+\langle F_\alpha\rangle,
\label{aux5}
\end{eqnarray}
is given by the gradient of the PMF,
\begin{eqnarray}
    \langle F_\alpha^{tot}\rangle=-\frac{\partial}{\partial X_\alpha} (V+\Delta V)=-\frac{\partial }{\partial X_\alpha}\, V_*(X). 
    \label{F_tot}
\end{eqnarray}
If the system is a single Brownian particle (as we assume in the sections to follow), the index $\alpha$ in the above equations is redundant and to be dropped.

We still need   to specify the parameter $c$ in definition (\ref{DV_def}) of the bath-induced potential  $\Delta V$.
In that expression, $c$ determines only an  additive constant (with respect to $X$) and, therefore, can be chosen arbitrarily. 
A common and often convenient choice is
\begin{eqnarray}
    c=Z_b,
\label{c1}
\end{eqnarray}
where $Z_b$ is the partition function for the equilibrium distribution of the
isolated (not connected to the system) bath, 
\begin{eqnarray}
    \rho_b=\frac{1}{Z_b}\, e^{-\beta H_b}, \quad Z_b=\Tr_b\left(e^{-\beta H_b}\right).
\end{eqnarray}
In  that case Eq. (\ref{DV_def}) gives for the bath-induced potential the following expression
\begin{eqnarray}
    \Delta V=-\frac{1}{\beta} \ln\left[
    \frac{Z_0}{Z_b}
    \right]=-\frac{1}{\beta}
    \ln\left[\frac{\Tr_b\left( e^{-\beta H_0}\right)}{\Tr_b\left( e^{-\beta H_b}\right)}
    \right],
    \label{DV}
\end{eqnarray}
and the HMF $H_s^*=H_s+\Delta V$ may be expressed as
\begin{eqnarray}
    H_s^*=-\frac{1}{\beta}
    \ln\left[\frac{\Tr_b\left( e^{-\beta H}\right)}{\Tr_b\left( e^{-\beta H_b}\right)}
    \right], 
    \label{HMF}
\end{eqnarray}
where $H=H_s+H_0$ is the total Hamiltonian of the combined system and the bath.

The advantage of the choice (\ref{c1}) can be seen considering the weak-coupling limit, 
when 
\begin{eqnarray}
H_0=H_b+\Phi\to H_b, \quad
H=H_s+H_0\to H_s+H_b,
\end{eqnarray}
and therefore
\begin{eqnarray}
Z_0(X)\to Z_b,\quad Z\to Z_s Z_b.
\label{Z_approx}
\end{eqnarray}
In that limit
the above equations 
give 
\begin{eqnarray}
    \Delta V\to 0,\quad V^*\to V,  \quad H_s^*\to H_s. 
    \label{limit}
\end{eqnarray}
That is a plausible result: For a negligible system-bath interaction
the bath-induced potential vanishes, and
the equilibrium statistics of the system is determined by the system's bare Hamiltonian $H_s$.

Let us note that expression (\ref{DV}) for $\Delta V$  often appears in the literature 
in a different form. Since
\begin{eqnarray}
    \frac{Z_0}{Z_b}=\frac{1}{Z_b}\, \Tr_b\left(
    e^{-\beta\,H_b}\,e^{-\beta \Phi}\right)=
    \langle e^{-\beta \Phi}\rangle_b,
\end{eqnarray}
where the average is taken over bath variables with distribution $\rho_b$, $\langle \cdots\rangle_b=\Tr_b(\cdots\rho_b)$, the expression (\ref{DV})
can be also written as
\begin{eqnarray}
    \Delta V=-\frac{1}{\beta}\, \ln
    \langle e^{-\beta \Phi}\rangle_b.
    \label{DV2}
\end{eqnarray}
Similarly, the HMF (\ref{HMF}) can be expressed as
\begin{eqnarray}
    H_s^*=-\frac{1}{\beta}\, \ln
    \langle e^{-\beta (H_s+\Phi)}\rangle_b.
    \label{HMF2}
\end{eqnarray}
Although Eqs. (\ref{DV2}) and (\ref{HMF2}) may be convenient in some cases~\cite{Talkner}, they will not appear in the rest of the paper.

Had we chosen the parameter $c$ in a form different than (\ref{c1}),  the bath-induced potential $\Delta V(X)$ would not vanish in the weak-coupling limit but instead reduce to a nonzero constant, 
\begin{eqnarray}
    \Delta V(X)=-\frac{1}{\beta} \ln\left[
    \frac{Z_0(X)}{c}\right]
    \to -\frac{1}{\beta} \ln\left[
    \frac{Z_b}{c}
    \right].
    \label{aux11}
\end{eqnarray}
That might be inconvenient,  yet it is not incorrect, since observable
physical quantities do not depend on $c$, provided $c$ is a constant with respect to $X$ and $P$.
In Ref.~\cite{Gelin}, the choice $c=1$ was adopted. It was criticized in~\cite{Talkner} on the ground that for such a choice $\Delta V$ does not vanish in the weak coupling limit.  In our view, 
 the only defect of the choice $c=1$ is that 
 the argument of the logarithm in Eq. (\ref{aux11}) is not dimensionless.

In what follows, we shall consider a model for which $H_0$ has the special form
\begin{eqnarray}
    H_0=h_1(x,X)+h_2(X), 
    \label{H0_special}
\end{eqnarray}
where $h_2(X)$ depends only on the system's position $X$, while $h_1(x,X)$ depends on  bath variables and $X$, yet the corresponding partition function
\begin{eqnarray}
    Z_1=\Tr_b\left(
    e^{-\beta h_1}\right)
\end{eqnarray}
does not depend on $X$. In that case, 
\begin{eqnarray}
Z_0(X)=Z_1\,e^{-\beta\, h_2(X)},    
\label{Z0_special}
\end{eqnarray}
and it is convenient to define  $c=Z_1$. Then, according to 
(\ref{DV_def}), the bath-induced potential is equal to $h_2(X)$,
\begin{eqnarray}
    \Delta V(X)=-\frac{1}{\beta} \ln\left[
    \frac{Z_0(X)}{Z_1}\right]=h_2(X).
    \label{DV_alt}
\end{eqnarray}

\section{Anchored bath model}
As discussed in the Introduction, 
the absence of a PMF in simplified models of Brownian motion is to some extent accidental:
If the system of interest is a single particle with no internal degrees of freedom and 
the bath is a translationally invariant fluid, then the average force from the bath on the system is zero by symmetry. Our goal here is to study a model where the system of interest is still a
structureless Brownian particle, but the translational invariance of the bath is broken, which results  in a non-trivial PMF. 
 
The bare Hamiltonian of the system is 
\begin{eqnarray}
    H_s=\frac{P^2}{2M}+V_{ex}(X),
\end{eqnarray}
where $V_{ex}(X)$ is the external potential unrelated to the bath.  
The bath  consists  of a large number $N$ of particles and has the bare Hamiltonian 
\begin{eqnarray}
    H_b=\sum_{i=1}^N 
    \frac{p_i^2}{2m}+\sum_{i>j}^N U(x_i-x_j)+\sum_{i=1}^N U_a(x_i).
    \label{Hb}
\end{eqnarray}
Here, the first two sums represent  the kinetic and pairwise potential energies.
The key feature of the model is the presence of the third sum, which implies that each particle in the bath is subject to the additional ``anchor" potential $U_a(x_i)$.  A particular realization of such a model, where the bath is an anchored harmonic chain, 
is sketched in Fig. 1. It is intuitively clear that the  anchored bath is non-passive and generates a PMF which tends to localize the system.
Physically, the model of the anchored   
bath may be relevant for  complex heterogeneous systems, like polymer networks and gels, characterized by a broad hierarchy of relaxation times~\cite{Milster}. In such systems, heavier (slower) structural units  may be essentially immobile on a relevant time scale and act 
as fixed anchoring sites for lighter and faster units.

We shall assume that the system-bath interaction energy $\Phi$ is pairwise,
\begin{eqnarray}
    \Phi=\sum_{i=1}^N\phi(X-x_i),
\end{eqnarray}
and write the total Hamiltonian in the form (\ref{H}),
\begin{eqnarray}
    H=H_s+H_0, \quad H_0(X)=H_b+\Phi(X),
    \label{H2}
\end{eqnarray}
where $H_0(X)$ is the Hamiltonian of the bath in the potential of the Brownian particle
held fixed  at position $X$.



The total Liouville operator for the model is convenient to write as
\begin{eqnarray}
    L=L_s+L_0
    \label{L}
\end{eqnarray}
where $L_s$ and $L_0$ correspond to Hamiltonians $H_s$ and $H_0$, respectively,
\begin{eqnarray}
    L_s=\frac{P}{M}\frac{\partial}{\partial X}+(F+F_{ex})\frac{\partial}{\partial P}, 
    \quad 
    L_0=\sum_i\left\{\frac{p_i}{m}\frac{\partial}{\partial x_i}+f_i\frac{\partial}{\partial p_i}\right\}.  
    \label{L2}
\end{eqnarray}
Here $F$ and $F_{ex}$ are  the forces on the particle due to the interaction with 
the bath and the external potential, respectively,
\begin{eqnarray}
    F=-\frac{\partial \Phi}{\partial X}, \quad 
    F_{ex}=-\frac{\partial V_{ex}}{\partial X},
    \label{forces}
\end{eqnarray}
and $f_i$ is the force on $i$-th particle of the bath 
\begin{eqnarray}
    f_i=-\frac{\partial H_0}{\partial x_i},
\end{eqnarray}
which parametrically depends on the system's position $X$.

We adopt the usual assumption of the Brownian motion theory that masses $m$ of bath particles are much smaller than the mass of the system, $m\ll M$.
In that case,  the bath is expected to evolve much faster than the system and 
quickly adjust to the instantaneous system's position.  Therefore, one 
 may expect that the bath's dynamics is close to that described by the Liouville operator $L_0$, which describes the bath in the field of a fixed (or, equivalently, an infinitely heavy) system.  The operator $L_0$ has the  properties
 \begin{eqnarray}
     L_0 H_0=0, \quad L_0 \,\varphi(H_0)=0,
     \label{aux40}
 \end{eqnarray}
where $\varphi(x)$ is an arbitrary differentiable function. A related  property is that $L_0$ and
$\varphi(H_0)$ commute,
\begin{eqnarray}
    L_0\,\varphi(H_0) A=\varphi(H_0)\, L_0 A,
    \label{aux42}
\end{eqnarray}
where $A$ is an arbitrary dynamical variable.
These properties of $L_0$ are generic and not affected  by the presence of the anchor potential. 


In the next two sections we shall use 
the Mazur-Oppenheim projection operator technique~\cite{MO,DO,KO}
to derive a Langevin equation with the PMF 
for the model described in this section.
The derivation is convenient to present in two steps. The first step is to re-write the equation of motion for the system separating the systematic and noisy components of the total force on the system. The result is what  can be called the pre-Langevin equation, see Eq. (\ref{pre_Langevin}) below. That equation is exact but impractical because it involves the system's variables in an implicit form.  The second step, discussed in Section 5, is to make that dependence explicit by applying an appropriate perturbation technique.

\section{Pre-Langevin equation}
The Mazur-Oppenhein technique~\cite{MO} is based on the projection operator $\mathcal P$ which averages an arbitrary dynamical variable $A$
over initial values of the bath variables
\begin{eqnarray}
    \mathcal{P} A=\langle A \rangle=\Tr_b({\rho_0\,A})
    \label{projector}
\end{eqnarray}
with distribution (\ref{rho_0})
\begin{eqnarray}
   \rho_0=\frac{1}{Z_0(X)}\, e^{-\beta H_0}, \quad 
    Z_0(X)=\Tr_b\left(e^{-\beta H_0}\right).
    \label{rho_02}
\end{eqnarray}
Recall that $\rho_0$ is the equilibrium distribution of the bath in the field of  the system fixed at $X$. This makes the projection (\ref{projector}) particularly relevant in the present context, because
$\rho_0$ is the distribution used in the definition of the average force related to the PMF, see Eqs. (\ref{F_av_def})-(\ref{F_tot}).

 The projection property $\mathcal P^2=\mathcal P$ is obvious, and  
another key property
\begin{eqnarray}
    \mathcal P L_0=0
\end{eqnarray}
follows from (\ref{aux42}) with $\varphi(H_0)=\rho_0$.
The complementary projection operator 
\begin{eqnarray}
    \mathcal Q=1-\mathcal P
\end{eqnarray}
has the properties
\begin{eqnarray}
    \mathcal P\mathcal Q=\mathcal Q\mathcal P=0, \quad \mathcal Q L_0=L_0.
    \label{property3}
\end{eqnarray}
The standard Mazur-Oppenheim approach 
assumes that  the bath is passive, i. e. 
\begin{eqnarray}
    \mathcal P F=\langle F\rangle=0,
\label{F_zero}
\end{eqnarray}
where $F$ is the force exerted by the bath on the Brownian particle. Here we consider an extension when Eq. (\ref{F_zero}) does not hold.

The equation of motion for the system reads 
\begin{eqnarray}
    \dot P(t)=F(t)+F_{ex}(t),
    \label{aux5000}
\end{eqnarray}
where $F$ is the force from the bath and $F_{ex}$ is the external force, see Eq. (\ref{forces}).
The time dependence of forces can be described using the Liouville propagator
\begin{eqnarray}
    F(t)=e^{L t}\, F, \quad F_{ex}(t)=e^{L t}\, F_{ex},
    \label{aux500}
\end{eqnarray}
where $F=F(0), F_{ex}=F_{ex}(0)$.  We shall also use a similar notation $A=A(0)$ for
the initial values of 
other dynamical variables.  
Inserting the identity operator $1=\mathcal Q+\mathcal P$ into expression (\ref{aux500}) for $F(t)$,
\begin{eqnarray}
    F(t)=e^{Lt} \mathcal P F+e^{Lt} \mathcal Q F
    =e^{Lt} \langle F\rangle+e^{Lt} \mathcal Q F
\end{eqnarray}
brings the equation of motion (\ref{aux5000}) to the form
\begin{eqnarray}
    \dot P(t)=e^{Lt}\left[
    \langle F\rangle+F_{ex}\right]+e^{Lt} \mathcal Q F.
    \label{aux53}
\end{eqnarray}
According to Eqs. (\ref{F_av}) and (\ref{F_tot}),
we can express $\langle F\rangle$ as the negative gradient of the the bath-induced potential $\Delta V$, 
\begin{eqnarray}
    \langle F\rangle=-\frac{\partial}{\partial X}\,\Delta V, 
\end{eqnarray}
and the total force $\langle F\rangle+F_{ex}$ as the negative gradient  of the PMF $V_*(X)$,  
\begin{eqnarray}
  \langle F\rangle+F_{ex}=- \frac{\partial}{\partial X}\, V_*(X), \qquad 
  V_*(X)=\Delta V(X)+V_{ex}(X).
\end{eqnarray}
Then the equation of motion (\ref{aux53}) can be written as
\begin{eqnarray}
    \dot P(t)=-\nabla V_*[X(t)]+e^{Lt} \mathcal Q F,
    \label{aux533}
\end{eqnarray}
where the first term on the right-hand side  is the gradient of the  PMF at time $t$,
\begin{eqnarray}
\nabla V_*[X(t)]=e^{Lt} \frac{\partial}{\partial X} V_*(X)=\frac{\partial}{\partial X} V_*(X)\Big{|}_{X=X(t)}.
\end{eqnarray}
That term is a conservative  part of the total force acting on the system.
The second term  $e^{Lt} Q F$ can be separated into a dissipative (non-conservative) force and noise.  
The usual way to facilitate such separation is to  use 
the operator identity 
\begin{eqnarray}
    e^{(A+B)t}=e^{At}+\int_0^t d\tau \,e^{A(t-\tau)} B\, e^{(A+B)\tau}. 
    \label{identity}
\end{eqnarray}
With $A=L$ and $B=-\mathcal P L$, one splits the propagator into two parts,
\begin{eqnarray}
    e^{Lt}=e^{\mathcal Q L t}+\int_0^t d\tau \,e^{L(t-\tau)}\, \mathcal P L  \,e^{\mathcal Q L \tau}.
\end{eqnarray}
Substituting  that into Eq. (\ref{aux533})  and recalling that $\mathcal P L_0=0$ yields
\begin{eqnarray}
    \dot P(t)=-\nabla V_*[X(t)]+F^\dagger(t)+\int_0^t d\tau\, e^{L(t-\tau)}\,\mathcal P L_s\,F^\dagger(\tau)
    \label{aux54}
\end{eqnarray}
where
\begin{eqnarray}
    F^\dagger(t)=e^{\mathcal Q L t} \mathcal Q F
\end{eqnarray}
is often referred to as the projected force.
Since $\mathcal P \mathcal Q=0$, the average $F^\dagger(t)$ is zero
\begin{eqnarray}
\langle F^\dagger(t)\rangle=\mathcal P F^\dagger(t)\sim \mathcal{P Q}=0,
\label{aux55}
\end{eqnarray}    
which suggests interpreting $F^\dagger(t)$ as noise. However, one has to keep in mind that  $F^\dagger(t)$ depends on the system's variables, and that dependence still needs to be revealed.

The integral term 
 in Eq. (\ref{aux54}) is   expected to give rise to a dissipation force.  Let us express 
$\mathcal P L_s F^\dagger(\tau)$ in the integrand in a more explicit form,
\begin{eqnarray}
    \mathcal P L_s F^\dagger(\tau)=
    \frac{P}{M}\, \left\langle
    \frac{\partial}{\partial X} F^\dagger(\tau)\right\rangle
    +\frac{\partial}{\partial P}\,\langle F\,F^\dagger(\tau)\rangle.
\label{aux50}
\end{eqnarray}
To evaluate the first term on the right-hand side, 
let us differentiate Eq. (\ref{aux55}), 
\begin{eqnarray}
    \frac{\partial}{\partial X} \left\langle F^\dagger(\tau)\right\rangle=
    -\beta \left\langle 
\frac{\partial H_0}{\partial X}\, F^\dagger(\tau)
\right\rangle
-\langle F^\dagger(\tau)\rangle \frac{\partial}{\partial X} \ln Z_0 +
\left\langle
    \frac{\partial}{\partial X} F^\dagger(\tau)\right\rangle=0.
\label{aux41}
\end{eqnarray}
Here we take into account that  for a non-passive bath the distribution 
$\rho_0=Z_0^{-1} \exp(-\beta H_0)$ depends on $X$ through both $H_0$ and $Z_0$.
However, the term with $\frac{\partial}{\partial X} \ln Z_0$  vanishes because of the factor $\langle F^\dagger(\tau)\rangle=0$.
As a result, one obtains from Eq. (\ref{aux41}) 
\begin{eqnarray}
\left\langle
    \frac{\partial}{\partial X} F^\dagger(\tau)\right\rangle=
    -\beta \left\langle
    F \, F^\dagger(\tau)\right\rangle,
\label{relation1}
\end{eqnarray}
and Eq. (\ref{aux50}) takes the form
\begin{eqnarray}
    \mathcal P L_s F^\dagger(\tau)=
    \left(\frac{\partial}{\partial P}- \frac{\beta\,P}{M}
    \right) \langle F \,F^\dagger(\tau)\rangle.
    \label{aux57}
\end{eqnarray}
Substituting into (\ref{aux54}) yields the equation of motion in the pre-Langevin form
\begin{eqnarray}
    \dot P(t)=-\nabla V_*[X(t)]+F^\dagger(t)+\int_0^t d\tau\, e^{L(t-\tau)}\,
     \left(\frac{\partial}{\partial P}- \frac{\beta\,P}{M}
    \right) \langle F \,F^\dagger(\tau)\rangle.
    \label{pre_Langevin}
\end{eqnarray}
This equation is similar to that for a passive bath~\cite{MO,KO},
with the only visible difference that  the external potential  $V_{ex}$ is replaced by the PMF $V_*=V_{ex}+\Delta V$. A hidden difference is that the projected force $F^\dagger(t)=e^{\mathcal Q L t} \mathcal Q F$ has the additional factor $\mathcal Q$, which ensures that $\langle F^\dagger(t)\rangle\sim \mathcal {PQ}=0$ for any time.
The equation is exact, but not yet operational because $F^\dagger(t)$ depends on the system's variables. That dependence can be made explicit   
by expanding the projected force in powers of a small mass ratio parameter $\lambda$, $F^\dagger(t)=F_0(t)+\lambda\, F_1(t)+\cdots$.
For a passive bath, in the leading (zeroth) order in $\lambda$ the projected force $F^\dagger(t)\approx F_0(t)$ does not depend on the system's variables, and Eq. (\ref{pre_Langevin}) is readily reduced to the standard generalized Langevin equation.  However, as we shall see in the next section, for a non-passive bath $F^\dagger(t)$ depends on the system's position even in the leading approximation. That dependence is a priori unknown, which may undermine the derivation of a practically useful Langevin equation with a PMF.

\section{Langevin equation}
 In order to bring the  pre-Langevin  Eq. (\ref{pre_Langevin}) to a more amiable form, let us  exploit a perturbation 
technique with the mass ratio
\begin{eqnarray}
    \lambda^2=m/M
\end{eqnarray}
as a small parameter.
If the overall system is not too far from equilibrium, the  momentum $P$ of the Brownian particle is close to
the equilibrium value $\sqrt{M/\beta}$, and the  
typical momentum of bath particles  
is $\sqrt{m/\beta}$. Therefore, the scaled momentum $P_*=\lambda P$ of the Brownian particle is on average of the same  order as momenta of the bath particles.
To make the $\lambda$-dependence
more explicit, let us re-write the pre-Langevin equation (\ref{pre_Langevin}) in terms of the scaled  momentum $P_*$,
\begin{eqnarray}
    \dot P_*(t)=-\lambda\,\nabla V_*[X(t)]+\lambda\,F^\dagger(t)+\lambda^2\int_0^t d\tau\, e^{L(t-\tau)}\,
     \left(\frac{\partial}{\partial P_*}- \frac{\beta\,P_*}{m}
    \right) \langle F \,F^\dagger(\tau)\rangle
    .
    \label{pre-Langevin}
\end{eqnarray}
The Liouville operator (\ref{L}) can be re-written as
\begin{eqnarray}
   L=L_0+\lambda L_1  
\end{eqnarray}
where $L_0$ is still given by Eq. (\ref{L2}) and 
\begin{eqnarray}
    L_1=\frac{P_*}{m}\frac{\partial}{\partial X}+(F+F_{ex})\frac{\partial}{\partial P_*}.
    \label{L3}
\end{eqnarray}
The projected force $F^\dagger(t)$ evolves as 
\begin{eqnarray}
    F^\dagger(t)=e^{\mathcal Q L t} \mathcal Q F=e^{(L_0+\lambda\mathcal Q L_1) t}\, \mathcal Q F,
    \label{aux51}
\end{eqnarray}
where we have used Eq. (\ref{property3}), $\mathcal QL_0=L_0$. Using
Eq. (\ref{identity}) with $A=L_0$ and $B=\lambda \mathcal Q L_1$ one finds
\begin{eqnarray}
e^{(L_0+\lambda\,\mathcal Q \,L_1)\, t}=e^{L_0t}+ 
\lambda\int_0^t d\tau e^{L_0\,(t-\tau)} \mathcal Q \,L_1 \,e^{L_0\tau}+
\mathcal O(\lambda^2).
\label{aux52}
\end{eqnarray}
Substituting into (\ref{aux51}) yields
\begin{eqnarray}
    F^\dagger(t)=E_0(t)+\lambda E_1(t)+\mathcal O(\lambda^2),
    \label{aux530}
\end{eqnarray}
where
\begin{eqnarray}
    E_0(t)=e^{L_0t} \mathcal Q F,\qquad
     E_1(t)=
    \frac{1}{m}\int_0^t d\tau\,P_*(t-\tau)\,e^{L_0(t-\tau)} \mathcal Q \frac{\partial}{\partial X}\,E_0(\tau),
    \label{aux531}
\end{eqnarray}
with both terms being zero-centered,
\begin{eqnarray}
    \langle E_0(t)\rangle=\langle E_1(t)\rangle\sim \mathcal {PQ}=0.
\end{eqnarray}
To order $\lambda^2$ the integral term in Eq. (\ref{pre-Langevin}) is simplified and the equation takes the form 
\begin{eqnarray}
    \dot P_*(t)=-\lambda\,\nabla V_*[X(t)]+\lambda \,F^\dagger(t)-\frac{\lambda^2\beta}{m}\int_0^t d\tau\, P_*(t-\tau)\, \, e^{L(t-\tau)} \langle F \,E_0(\tau)\rangle. 
    \label{gle3}
\end{eqnarray}
Since this equation is an approximation  to order $\lambda^2$, it can only be used to evaluate properties which are either linear or quadratic in noise $\lambda\,F^\dagger(t)$. 
Upon averaging, the contributions linear in noise vanish, while contributions quadratic in noise are determined to order $\lambda^2$ by the leading term $E_0(t)$ in the expansion (\ref{aux530}).
Therefore, perturbation consistency requires to  replace in Eq. (\ref{gle3}) the projected force $F^\dagger(t)$ by the leading-order approximation $E_0(t)$.  After that replacement, and after returning from the scaled to  the true momentum $P=P_*/\lambda$, Eq. (\ref{gle33}) takes the form 
\begin{eqnarray}
    \dot P(t)=-\nabla V_*[X(t)]+E_0(t)-\frac{\beta}{M}\int_0^t d\tau\, P(t-\tau)\, \, e^{L(t-\tau)} \langle F \,E_0(\tau)\rangle. 
    \label{gle33}
\end{eqnarray}
The fluctuating force $E_0(t)$ can be written as 
\begin{eqnarray}
    E_0(t)=F_0(t)-\langle F\rangle,\quad  F_0(t)=e^{L_0 t} F
\end{eqnarray}
and interpreted as the deviation of the force $F_0(t)$, exerted by the bath on the system fixed at $X$,  from the average value $\langle F(X)\rangle$. 
Clearly,  $E_0(t)$ does not depend on $P$, but  it does depend on the system's position $X$.

That dependence does not complicate the matter if  the bath is a homogeneous non-anchored fluid. In that case,  the dependence of forces on positions occurs only through the differences $X-x_i$ and $x_i-x_j$. When taking average over the bath coordinates $x_i$, one can change the integration variables $x_i'=x_i-X$, so that the 
integrands depend on $x_i'$ and $x_i'-x_j'$, but not on $X$. Therefore, although $F$ and $F_0(t)$ depend on $X$, their moments and correlations do not.  
Moreover, $\langle F\rangle=0$ by symmetry.
As a result, the PMF is reduced to the external potential, $\nabla V_*=\nabla V_{ex}$, $E_0(t)$ coincides with $F_0(t)$, and  the correlation   
$\langle F\, E_0(t)\rangle=\langle F\,F_0(t)\rangle$ does not depend on $X$.
Clearly, that correlation  also does not depend on $P$ and bath variables. Therefore, the propagator $e^{Lt}$ in the integral term of Eq. (\ref{gle33}) can be dropped, and the equation takes the form of the standard generalized Langevin equation
\begin{eqnarray}
    \dot P(t)= -\nabla V_{ex}[X(t)]+F_0(t)- \frac{\beta}{M}\int_0^t d\tau\, P(t-\tau)\, \langle F \,F_0(\tau)\rangle,
\end{eqnarray}
supplemented with the ordinary fluctuation-dissipation relation.

For an anchored bath, the situation is more complicated
because the statistical properties of noise $E_0(t)$ generally depend on $X$. While
 forces $F$ and $F_0(t)$ still depend on $X$ only through $X-x_i$, the internal bath forces 
depend not only on coordinate differences  $x_i-x_j$, but also on the positions of individual bath units $x_i$. In that case, 
the variable change $x_i'=x_i-X$ does not eliminate $X$ from the averages. As a result,  
$\langle F\rangle$ and $\langle F F_0(t)\rangle$ depend on $X$ and so does the correlation 
\begin{eqnarray}
    \langle F E_0(t)\rangle=\langle F F_0(t)\rangle-\langle F\rangle^2. 
\end{eqnarray}
One can reflect that dependence denoting
\begin{eqnarray}
   \frac{\beta}{M}\langle F E_0(t)\rangle=\Gamma(t,X).
    \label{fdr2}
\end{eqnarray}
Then  Eq. (\ref{gle33}) takes the form
\begin{eqnarray}
    \dot P(t)=-\nabla V_*[X(t)]-E_0(t)(t)-\int_0^t d\tau\, P(t-\tau)\, \Gamma[\tau, X(t-\tau)],
    \label{gle4}
\end{eqnarray}
which is formally similar to the  standard  generalized Langevin equation (with the external potential 
replaced by the PMF), except that  
the dissipation term is not of the conventional convolution form since it depends on both $\tau$ and $t-\tau$. As it is, the equation is not operational because the dependence  of the kernel on $X$ is a priory unknown. That is not necessarily a problem.
After all,  the exact form of the PMF $V_*(X)$ is usually not known either; it is often defined empirically to fit experimental data. One may define the function $\Gamma(X)$ empirically as well. However, the difficulty is  that the two functions $V_*(X)$ and $\Gamma(X)$ are clearly not independent, yet there seems to be no general relation describing their connection.

The situation is simplified when the Hamiltonian $H_0$ has a special structure (\ref{H0_special}), 
$H_0=h_1(x,X)+h_2(X)$, and therefore 
the force $F_0(t)$ on the fixed system has the form
\begin{eqnarray}
    F_0(t)=e^{L_0 t} \left(-\frac{\partial H_0}{\partial X}\right)=
    f_1(X,x,t)+f_2(X),
    \label{special}
\end{eqnarray}
where component $f_2(X)$ depends only on the system's position. In addition, let us assume that  component $f_1$ has the properties of noise  in the standard theory of Brownian motion. Namely,  the average of $f_1$ over bath variables (with distribution $\rho_0$) is zero for any $t$, 
\begin{eqnarray}
    \langle f_1(X,x,t)\rangle=0,
\label{special2}
\end{eqnarray} 
and its auto-correlation correlation 
\begin{eqnarray}
    \langle f_1(X,x,0)\,f_1(X,x,t)\rangle\equiv \langle f_1 f_1(t)\rangle=c(t)
    \label{special3}
\end{eqnarray}
does not depend on the system's position $X$. 
In that case, 
\begin{eqnarray}
    \langle F\rangle=f_2(X), \quad
    E_0(t)=F_0(t)-\langle F\rangle=f_1(X,x,t),
\end{eqnarray}
and the correlation  $\langle F\,E_0(t)\rangle$ coincides with the auto-correlation of $f_1$
\begin{eqnarray}
    \langle F E_0(t)\rangle=\langle (f_1+f_2)\, f_1(t)\rangle=\langle f_1 f_1(t)\rangle+f_2\langle f_1(t)\rangle=
    \langle f_1 f_1(t)\rangle,
\end{eqnarray}
and therefore, because of assumption (\ref{special3}), does not depend on $X$.   
Then in Eq. (\ref{gle3})
the propagator $e^{L(t-\tau)}$ has no effect and can be dropped. Also, since noise $E_0(t)$ is zero-centered,   the correlation $\langle F E_0(t) \rangle $ coincides with the auto-correlation of $E_0(t)$, 
\begin{eqnarray}
    \langle E_0(0)\,E_0(t)\rangle=\langle F\,E_0(t)\rangle-\langle F\rangle\,\langle E_0(t)\rangle=\langle F\,E_0(t)\rangle.
\end{eqnarray}
As a result, Eq. (\ref{gle3}), written in terms of the true momentum $P=P_*/\lambda$,
turns into 
the standard generalized Langevin equation
\begin{eqnarray}
    \dot P(t)= -\nabla V_{ex}[X(t)]+E_0(t)-\frac{\beta}{M}\int_0^t d\tau\, P(t-\tau)\, \langle E_0(0) \,E_0(\tau)\rangle
    \label{gle6}
\end{eqnarray}
with the external potential $V_{ex}(X)$ replaced by the PMF $V_*(X)$.
This is the equation advocated in Refs.~\cite{Lange,Kinjo}.  We observe that the equation is valid  only under special assumptions. In the next two sections we consider a special (linear)  version of the model with an anchored bath where assumptions (\ref{special})-(\ref{special3}) do hold and therefore  the Langevin equation (\ref{gle6}) is valid.

\section {Klein-Gordon chain: the PMF}
Let us now consider a more specific model where the bath is formed by the  Klein-Gordon harmonic chain. The chain consists of
$N\gg 1 $ atoms of mass $m$, connected by springs with stiffness $k_0$.  Each atom of the chain 
is subjected to the on-site anchor harmonic potential with stiffness 
$k_a$. The system of interest is a particle of mass $M$, connected by a $k_0$-spring to the first atom of the chain and subjected to external potential $V_{ex}(X)$. We assume a boundary condition when 
the terminal atom of the chain is connected by a $k_0$-spring to a wall.
The model is sketched in Fig. 1 and can be viewed as the familiar Rubin model~\cite{Zwanzig,Weiss}, extended by the presence of anchor potentials
acting on atoms of the bath and external potential $V_{ex}(X)$ acting on the system. It is well known that a non-anchored chain can
serve, in the limit $N\to\infty$, as a passive bath~\cite{Zwanzig,Weiss}. In contrast, 
the Klein-Gordon chain is non-passive: in thermal equilibrium it produces a systematic harmonic potential acting on the system, so the PMF has the form 
\begin{eqnarray}
    V_*(X)=V_{ex}(X)+\frac{1}{2}\,\kappa\, X^2.
\label{KG_pmf}
\end{eqnarray}
Below we evaluate stiffness $\kappa$ of the bath-induced potential;  the result is given by Eq. (\ref{kappa}).

We write the total Hamiltonian of the model in form (\ref{H2}),
\begin{eqnarray}
    H=H_s+H_0(X),
    \label{H22}
\end{eqnarray}
where $H_s$ is the Hamiltonian of the isolated system  
\begin{eqnarray}
    H_s=\frac{P^2}{2M}+ V_{ex}(X), 
\label{H222}
\end{eqnarray}
and $H_0(X)$ is the Hamiltonian of the bath interacting with the system fixed at position $X$,
\begin{eqnarray}
    H_0(X)=\sum_{i=1}^N \left\{
    \frac{p_i^2}{2m}+\frac{k_0}{2} (x_i-x_{i+1})^2+\frac{k_a}{2}\,x_i^2
    \right\}+\frac{k_0}{2}\, (X-x_1)^2.
\label{H00}
\end{eqnarray}
Coordinates $x_i$ represent displacements of the chain's atoms from positions of mechanical equilibrium. 
In accordance with the chosen boundary condition, we assume in Eq. (\ref{H00}) that $x_{N+1}=0$. 
It is convenient to present $H_0$ as
\begin{gather}
    H_0=H_0'+H_0'',\nonumber\\
    H_0'=\sum_{i=0}^N \left\{
    \frac{p_i^2}{2m}+\frac{k_0}{2} (x_i-x_{i+1})^2+\frac{k_a}{2}\,x_i^2
    \right\},\qquad 
    H_0''=-k_0\,x_1\,X+\frac{k_0}{2}\,X^2,
    \label{H000}
\end{gather}
assuming that $x_0=x_{N+1}=0$ and $p_0=0$.
Hamiltonian $H_0'$ describes the isolated Klein-Gordon chain of $N$ atoms with terminal atoms $i=1$ and $i=N$ connected to the walls, see Fig. 2.

\begin{figure}
\includegraphics[height=3.5cm]{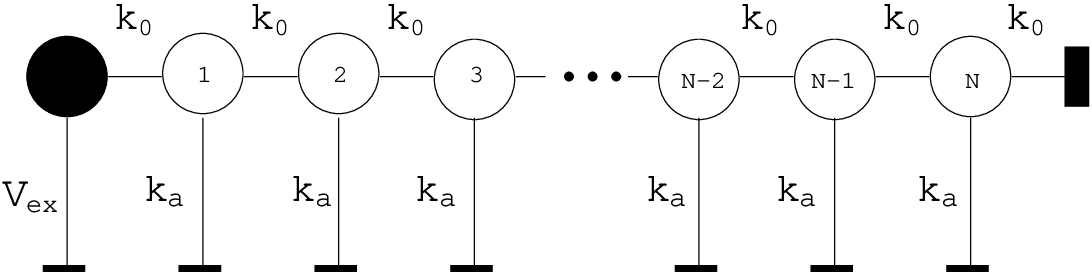}
\caption{The setup considered in Sections 6 and 7. The black circle represents the system of interest, which is a classical particle subjected to an (arbitrary) external potential $V_{ex}(X)$. The system is connected by the spring with stiffness $k_0$ to the thermal bath, formed by the Klein-Gordon lattice. The latter consists of $N$ atoms connected by the springs with stiffness $k_0$. Each atom of the lattice is subjected to on-site (anchoring) harmonic potentials with stiffness $k_a$. 
}
\end{figure}

The advantage of working with $H_0'$ is that it can be readily
diagonalized  using the well-known normal mode transformation $\{x_i,p_i\}\to \{Q_j,P_j\} $,
\begin{eqnarray}
x_i=\frac{1}{\sqrt{m}}\, \sum_{j=1}^N A_{ij}\,Q_j, \quad 
p_i=\sqrt{m} \,\sum_{j=1}^N A_{ij}\,P_j,    
\label{transformation}
\end{eqnarray}
where the transformation matrix
\begin{eqnarray}
    A_{ij} =\sqrt{\frac{2}{N+1}}\, \sin\frac{i j \pi}{N+1}
\end{eqnarray}
satisfies  the orthogonality relations 
\begin{eqnarray}
    \sum_{i=1}^N A_{ij}A_{ij'}=\delta _{jj'},\quad
\sum_{j=1}^N A_{ij}A_{i'j}=\delta _{i i'}.
\end{eqnarray}
The transformation has the same form as
for the corresponding 
harmonic chain with no anchor potentials. It is canonical and preserves phase-space volumes, i. e. the Jacobian of the transformation equals $1$.  Therefore, the average
$\langle\cdots \rangle $ over the bath variables $\{x_i,p_i\}$, which we used in the previous sections, can be replaced by  the average over $\{Q_j, P_j\}$.

Expressed in terms of  $\{Q_j,P_j\}$, Hamiltonians  $H_0'$  and $H_0''$ take the forms
\begin{eqnarray}
H_0'=\frac{1}{2} \sum_{j=1}^N \left\{
    P_j^2+\Omega_j^2\,Q_j^2\right\}, \qquad
H_0''=-\sum_{j=1}^N c_j\, Q_j\, X+
    \frac{k_0}{2}\,X^2,  
\label{H0_diag}
\end{eqnarray}
where the coupling coefficients $c_j$ are  
\begin{eqnarray}
    c_j=\frac{k_0}{\sqrt{m}} A_{1j}=
    \frac{k_0}{\sqrt{m}}
    \sqrt{\frac{2}{N+1}}\,\sin\frac{\pi j}{N+1},
\label{cj}
\end{eqnarray}
and the normal mode frequencies $\Omega_j$ are determined by the following expressions
\begin{eqnarray}
    \Omega_j^2&=&\omega_j^2+\omega_a^2, \quad \omega_j=\omega_0\sin\frac{\pi j}{2(N+1)},\quad
     \omega_0^2=4\,k_0/m, \quad \omega_a^2=k_a/m.
    \label{Omegaj}
\end{eqnarray}
In these expressions, $\omega_j$ are normal mode frequencies and $\omega_0$ is the highest normal mode frequency
in the unanchored chain (with $\omega_a=0$).
Note that for the anchored chain the spectrum of normal modes is bounded not only from above (as for the non-anchored chain), but also from below
\begin{eqnarray}
    \omega_a^2<\Omega_j^2<\omega_0^2+\omega_a^2.
\end{eqnarray}
This feature may be important
for ergodic properties of the combined system~\cite{Plyukhin_NEO}.

\begin{figure}
\includegraphics[height=3.5cm]{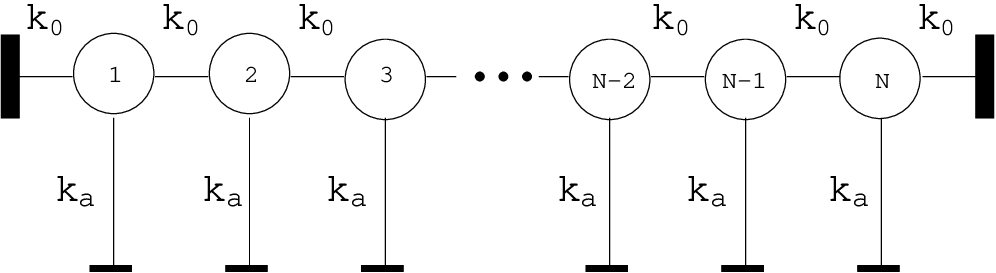}
\caption{The Klein-Gordon lattice described by Hamiltonian $H_0'$ in Eq. (\ref{H000}).}
\end{figure}

With $H_0'$ and $H_0''$ given by  Eq. (\ref{H0_diag}), Hamiltonian $H_0=H_0'+H_0''$ takes the form
\begin{eqnarray}
    H_0=\frac{1}{2}\sum_{j=1}^N\left\{
    P_j^2+\Omega_j^2\left[
    Q_j^2-\frac{2c_j}{\Omega_j^2}\, Q_j\,X\right]\right\}+\frac{k_0}{2}\,X^2.
    \label{H0_modes}
\end{eqnarray}
Completing the square in $[\cdots]$, one gets
\begin{eqnarray}
    H_0=\frac{1}{2}\sum_{j=1}^N\left\{
    P_j^2+\Omega_j^2\left[
    Q_j-\frac{c_j}{\Omega_j^2}\,X\right]^2\right\}+\frac{1}{2}\,
    \left[k_0-\sum_{j=1}^N\left(\frac{c_j}{\Omega_j}\right)^2\right]
    X^2.
\end{eqnarray}
In the limit $N\to\infty$ one finds
\begin{eqnarray}
    \sum_{j=1}^N \left(\frac{c_j}{\Omega_j}\right)^2\to 
    \frac{k_0}{\pi}\,\int_0^{\pi/2}
    \frac{\sin^2 (2x)\,  dx}{a^2+\sin^2x}=k_0(1+2 a^2-2 a\sqrt{1+a^2}),
\label{aux1}
\end{eqnarray}
where we introduced the parameter
$a=\omega_a/\omega_0$,
characterizing the relative strength of the anchor potential. 
Then, in the limit of the infinite chain, $H_0$ takes the form
\begin{eqnarray}
    H_0=\frac{1}{2}\sum_{j=1}^N\left\{
    P_j^2+\Omega_j^2\left[
    Q_j-\frac{c_j}{\Omega_j^2}\,X\right]^2\right\}+\frac{1}{2}\,
    \kappa\, X^2,
\end{eqnarray}
with
\begin{eqnarray}
    \kappa=2\,k_0\,a\left\{
    \sqrt{1+a^2}-a\right\}, \quad  a=\frac{\omega_a}{\omega_0}=\sqrt{\frac{k_a}{4\,k_0}}.
\label{kappa}
\end{eqnarray}

We observe that $H_0$ has the 
form (\ref{H0_special}), discussed at the end of Section 2, with  positions of the bath particles $x=\{x_i\}$ replaced by normal modes $Q=\{Q_j\}$,
\begin{gather}
    H_0=h_1(Q,X)+h_2(X), \nonumber\\
    h_1=\frac{1}{2}\sum_j\left\{
    P_j^2+\Omega_j^2\left[
    Q_j-\frac{c_j}{\Omega_j^2}\,X\right]^2\right\},\qquad 
    h_2(X)=\frac{1}{2}\,
    \kappa\, X^2.
    \label{H0_diag2}
\end{gather}
Using the substitution $Q'_j=Q_j-\frac{c_j}{\Omega_j^2}\,X$, one observes that  the partition function corresponding to $h_1$ 
\begin{eqnarray}
    Z_1=\Tr_b\left(e^{-\beta h_1}\right)
\end{eqnarray}
does not depend on $X$. Then the partition function $Z_0$ corresponding to $H_0$ has the form (\ref{Z0_special}),
\begin{eqnarray}
Z_0(X)=\Tr_b\left(e^{-\beta H_0}\right)=Z_1\,e^{-\beta\, h_2(X)},    
\end{eqnarray}
Defining the constant $c$ in the expression (\ref{DV_def}) of the bath-induced potential $\Delta V$ to be $c=Z_1$, we find that
\begin{eqnarray}
    \Delta V=-\frac{1}{\beta} \ln\left[
    \frac{Z_0(X)}{c}\right]=h_2(X)=\frac{1}{2}\,\kappa \,X^2,
\end{eqnarray}
and the PMF $V_*=V_{ex}+\Delta V$ has the form (\ref{KG_pmf}).

Thus, the bath formed by the infinite Klein-Gordon chain is non-passive; it acts on the system with the systematic average  force $\langle F\rangle=-\kappa\,X$.
The effective stiffness $\kappa$ of that force as a function of $a=\omega_a/\omega_0$ is given by Eq. (\ref{kappa}). It monotonically increases from $0$ at $a=0$ to $k_0$ for $a\gg 1$.
Both limiting values are intuitively plausible. For $a=0$, the model is equivalent to the Rubin model~\cite{Zwanzig,Weiss} where the chain is not anchored and in the limit $N\gg 1$ 
(but not for finite $N$)  
serves as a passive bath, which does not produce a systematic force on the fixed system. 
For $a\gg 1$ the chain acts as an infinitely heavy anchor.  In that case, the system feels only  the spring of stiffness $k_0$ by which it is attached to the terminal atom  of the chain.


\section{Klein-Gordon chain: Langevin equation}

We have seen at the end of Section 5 that a single-particle system subjected to a PMF is described (in the leading order in $\lambda$) by the standard generalized Langevin equation (\ref{gle6}) if statistical properties of  noise
\begin{eqnarray}
    E_0(t) =F_0(t)-\langle F\rangle,
    \label{aux70}
\end{eqnarray}
specifically the correlation $\langle E_0(0)\,E_0(t)\rangle$,
do not depend on the system's variables. Let us show that this is just the case for the model described in the previous section, i.e. when the bath, linearly coupled to the system, is formed by the harmonic Klein-Gordon chain. 

The microscopic force exerted by the bath on  the system $F=-k_0(X-x_1)$ can be written in terms of normal modes as follows 
\begin{eqnarray}
    F=-\frac{\partial H_0}{\partial X}=-\frac{\partial H_0''}{\partial X}=-k_0 X+\sum_j c_j Q_j,
    \label{F_aux1}
\end{eqnarray}
where $c_j$ are given by Eq. (\ref{cj}). Then $F_0(t)=e^{L_0 t}F$ has the form
\begin{eqnarray}
    F_0(t)=-k_0 X+\sum_j c_j Q_j(t). 
\label{F0_aux1}
\end{eqnarray}
The equations of motion for $Q_j(t)$, governed  by Hamiltonian $H_0$, Eq. (\ref{H0_modes}), read
\begin{eqnarray}
 \ddot Q_j(t)+\Omega_j^2 Q_j(t)=c_j X,   
\label{eq_motion}
\end{eqnarray}
where the system's  position $X$ is a fixed parameter. Solutions of these equations can be written as 
\begin{eqnarray}
    Q_j(t)=\left(Q_j-\frac{c_j}{\Omega_j^2}\, X\right)\,\cos\Omega_j t+\frac{1}{\Omega_j} P_j\,\sin \Omega_jt +\frac{c_j}{\Omega_j^2} X,
\end{eqnarray}
where $Q_j=Q_j(0)$ and $P_j=\dot Q_j(0)$.
Substituting to Eq. (\ref{F0_aux1}) yields
\begin{eqnarray}
    F_0(t)=\sum_j c_j\left\{\left(Q_j-\frac{c_j}{\Omega_j^2}\, X\right)
    \,\cos\Omega_j t+\frac{1}{\Omega_j} P_j\,\sin \Omega_jt\right\}-\kappa X
\label{aux71}
\end{eqnarray}
with
\begin{eqnarray}
    \kappa=k_0-\sum_{j=1}^N \left(
    \frac{c_j}{\Omega_j}
    \right)^2.
\end{eqnarray}
As was shown in the previous section, in the limit of the infinite chain, $N\to\infty$, the effective stiffness $\kappa$ takes the form  (\ref{kappa}).



We observe that $F_0(t)$ satisfies conditions (\ref{special})-(\ref{special3}),
mentioned at the end of Section 5, which ensure that
statistical properties of the zero-centered noise
$E_0(t)=F_0(t)-\langle F\rangle$ do not depend on $X$. We can also show that directly.
Indeed, 
using the substitution 
\begin{eqnarray}
  Q_j'=Q_j-\frac{c_j}{\Omega_j^2}\,X, \quad P_j'=P_j,
  \label{Q_new}
\end{eqnarray}
we observe that in terms of $Q_j', P_j'$ the Hamiltonian $H_0$ has the quadratic form
\begin{eqnarray}
    H_0=\frac{1}{2}\sum_j\left\{
    {P_j'}^2+\Omega_j^2\,{Q_j'}^2\right\}+\frac{1}{2}\,
    \kappa\, X^2,
    \label{H0_quad}
\end{eqnarray}
so that 
\begin{eqnarray}
    \langle Q'_j\rangle=0, \quad 
    \langle Q'_j Q'_k\rangle=
    \frac{\delta_{jk}}{\beta\,\Omega_j^2}, \quad 
    \langle P'_j P'_k\rangle=\frac{\delta_{jk}}{\beta}.
    \label{averages}
\end{eqnarray}
From these relations and Eq. (\ref{aux71})
one obtains
\begin{eqnarray}
    \langle F\rangle=-\kappa X.
\label{aux72}
\end{eqnarray}
Therefore, the zero-centered noise $E_0(t)=F_0(t)-\langle F\rangle$
has the form
\begin{eqnarray}
    E_0(t)=\sum_j c_j\left\{
    Q'_j\,\cos\Omega_j t+\Omega_j^{-1} P_j'\,\sin \Omega_jt\right\},
\label{noise}
\end{eqnarray}
and its correlation
\begin{eqnarray}
    \langle E\,E_0(t)\rangle=\frac{1}{\beta}\,\sum_{j=1}^N \left(
    \frac{c_j}{\Omega_j}\right)^2 
    \cos \Omega_j t
    \label{corr_special}
\end{eqnarray}
does not depend on $X$.
In that case, as discussed in Section 5, 
the Langevin equation takes the standard form (\ref{gle6}), which we can write as
\begin{eqnarray}
    \dot P(t)=-\nabla V_*[X(t)]+E_0(t)-\int_0^t d\tau\, P(t-\tau)\, \Gamma(\tau),
    \label{gle7}
\end{eqnarray}
where the PMF equals 
\begin{eqnarray}
    V_*(X)=\frac{1}{2}\,\kappa X^2+V_{ex}(X),
\end{eqnarray}
stiffness $\kappa$ of the bath-induced potential is given by Eq. (\ref{kappa}), 
and the dissipative kernel $\Gamma(t)$ is related to the zero-centered noise $E_0(t)$  by the standard fluctuation-dissipation relation 
\begin{eqnarray}
    \Gamma(t)=\frac{\beta}{M}\,\langle E E_0(t)\rangle.
\label{fdr7}
\end{eqnarray}

Recall that we derived the Langevin equation (\ref{gle7}) expanding the  projected force in powers of $\lambda=\sqrt{m/M}$ 
\begin{eqnarray}
    F^\dagger(t)=e^{\mathcal Q L t}F=E_0(t)+\lambda E_1(t)+\cdots
    \label{F_exp}
\end{eqnarray}
and restricting to the leading 
approximation 
$F^\dagger(t)\approx E_0(t)$. 
One can show that for the Klein-Gordon model the next term $E_1(t)$, which is given by  Eq. (\ref{aux531}), vanishes.
Indeed, as follows from Eqs. (\ref{Q_new}) and (\ref{noise}), $E_0(t)$ is linear in $X$, the derivative  
\begin{eqnarray}
    \frac{\partial E_0(t)}{\partial X}=\sum_j c_j\,\frac{\partial Q_j'}{\partial X}=-\sum_j\left(
    \frac{c_j}{\Omega_j}\right)^2
\end{eqnarray}
does not depend on bath variables, and therefore
\begin{eqnarray}
    E_1(t)\sim \mathcal Q\,\ \frac{\partial E_0(t)}{\partial X}=0.
\end{eqnarray}
In a similar way one can show that all terms of higher order in the expansion (\ref{F_exp}) vanish as well. Thus,   for the presented model  the projected force equals $E_0(t)$ exactly,
\begin{eqnarray}
    F^\dagger(t)=E_0(t),
\end{eqnarray}
and the Langevin equation (\ref{gle7}) is also exact.

\begin{figure}
\includegraphics[height=7.5cm]{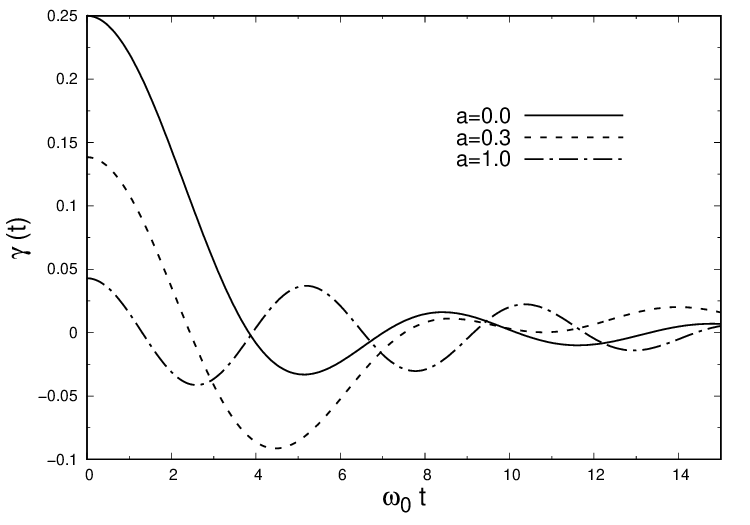}
\caption{ The dimensionless dissipation kernel $\gamma(t)$, see Eq. (\ref{aux75}), for three values of the anchor parameter $a=\omega_a/\omega_0$.
}
\end{figure}

The derivation produced above is, of course, not the most economical. 
We used the projection operator technique with the hope (unfulfilled) to get a Langevin equation for a general setting described in Section 3. Had we assumed the linearity of forces from the start, we could have derived the Langevin equation (\ref{gle7}) more easily
by the explicit integration method, i. e.  by integrating linear equation of motion of the bath~\cite{Zwanzig,Weiss,Hanggi}. We address that method in the Appendix.

We finish this section by deriving a more explicit expression for the correlation $\langle E E_0(t)\rangle$ and the dissipation kernel $\Gamma(t)$. According to 
Eqs. (\ref{corr_special}) and (\ref{fdr7}), the kernel equals
\begin{eqnarray}
    \Gamma(t)=\frac{1}{M}\,\sum_{j=1}^N \left(
    \frac{c_j}{\Omega_j}\right)^2 
    \cos \Omega_j t.
\label{Gamma}
\end{eqnarray}
Taking into account Eqs. (\ref{cj}) and (\ref{Omegaj}) for $c_j$ and $\Omega_j$, and taking 
the limit $N\to\infty$ we obtain 
\begin{eqnarray}
    \Gamma(t)=\lambda^2 \omega_0^2\,\gamma(t), \qquad \gamma(t)=\frac{1}{4\pi}\, \int_0^{\pi/2}
    \frac{\sin^2(2x)}{a^2+\sin^2x}\, \cos\left(\omega_0t \sqrt{a^2+\sin^2 x}\right)\,dx,
\label{aux75}
\end{eqnarray}
and the noise correlation equals
$\langle E E_0(t)\rangle=\frac{M}{\beta}\, \Gamma(t)$. 
For the non-anchored chain  ($a=0$), 
the dimensionless kernel $\gamma(t)$ is expressed in terms of the Bessel function $J_1(x)$,
\begin{eqnarray}
    \gamma(t)=
\frac1{2\,\omega_0\,t}\, J_1(\omega_0 t).
\end{eqnarray}
This is the well-known result for the Rubin model~\cite{Zwanzig,Weiss}. For  nonzero $a$ the kernel (\ref{aux75}) cannot be evaluated analytically. Fig. 3 presents numerical evaluations of $\gamma(t)$ for several values of $a$. 

\section{Conclusion}
The concept of the PMF naturally emerges in equilibrium statistical mechanics of complex systems consisting of many particles. The force $F_i$ on $i$-th  particle of the system is produced by the bath disturbed by the interaction with other parts of the system.  As a result, even if the bath is a translationally-invariant fluid, the mean force $\langle F_i\rangle$ is generally not zero and is determined by the bath-induced part of the PMF, denoted above as $\Delta V$. 
With the PMF it is 
straightforward to evaluate average values in equilibrium, but 
there is a variety of views on how to incorporate the PMF in the Langevin dynamics.

In this paper, we have simplified the problem by assuming that the system of interest consists of a single particle. In that case, the PMF is usually irrelevant because  the homogeneous bath is passive and produces no systematic force on a fixed system. That is not so, however, if the bath is not translationally-invariant. In the presented model the symmetry of the bath is broken by on-site anchor potentials applied to individual particles of the bath. 
Although that setting offers a simpler exposition, we were unable to derive a Langevin equation in a closed form. The non-passivity of the bath  not only modifies the average potential acting on the system, but also makes the dissipation kernel
and noise statistics depend on the position of the system $X$. 
That dependence is a priory unknown and there seems to be no general relation connecting it to the PMF.  
Therefore,  we argue that there is no general form for a Langevin equation with the PMF.  
The conclusion corroborates the findings of Refs.~\cite{Schilling,Schilling_review,Vroy}, although we used here a different projection operator technique.

An exception is the case where all internal forces in the combined system are linear. As a specific example, we considered the bath formed by the harmonic  Klein-Gordon chain. In that case, the dissipation kernel and noise statistics do not depend on $X$, and 
the Langevin equation has the standard form with the external potential replaced by 
the PMF. Such an equation  is often exploited in empirical studies, but we observe that it holds only under rather exceptional conditions.

\renewcommand{\theequation}{A\arabic{equation}}
  \setcounter{equation}{0}  

\section*{Appendix} 
In Section 7 we derived the Langevin equation for the model where the bath is represented by the harmonic Klein-Gordon chain by applying the generic method based on the projection operator technique. In this Appendix we present an alternative method based on explicit integration of equations of motion of the bath. 
Of course, that method is only feasible when the equations of motion 
are linear. The derivation is similar to those for 
the Caldeira-Leggett and Rubin models~\cite{Zwanzig,Weiss,Hanggi}. 

The Hamiltonian $H$ of the model is given by Eqs. (\ref{H22})-(\ref{H00}), and 
the equation of motion for the system is
\begin{eqnarray}
    \dot P=-\frac{\partial H}{\partial X}=-k_0\, (X-x_1)-\frac{\partial V_{ex}}{\partial X}.
\end{eqnarray}
Using transformation (\ref{transformation}) to the bath's normal modes, the equation can be written as
\begin{eqnarray}
    \dot P=-k_0\,X-\sum_j c_j Q_j-\frac{\partial V_{ex}}{\partial X},
\label{a_eom1}
\end{eqnarray}
where $c_j$ are given by Eq. (\ref{cj}).
The equations of motion for the bath's normal modes are
\begin{eqnarray}
    \dot P_j=-\frac{\partial H}{\partial Q_j}=-\Omega_j^2 Q_j+c_j X,\quad
    \dot Q_j=\frac{\partial H}{\partial P_j}=P_j, 
\end{eqnarray}
Excluding $P_j$, one obtains 
\begin{eqnarray}
    \ddot Q_j(t)+\Omega_j^2\, Q_j(t)=c_j X(t).
\label{a_eom2}
\end{eqnarray}
The general solution  has the form
\begin{eqnarray}
   Q_j(t)=Q_j^0(t)+\frac{c_j}{\Omega_j}\,\int_0^t \sin (\Omega_j \tau)\,X(t-\tau)\,d\tau, 
   \label{a_sol1}
\end{eqnarray}
where $Q_j^0(t)$ 
is the general solution of the homogeneous equation $\ddot Q_j+\Omega_j^2\,Q_j=0$, 
\begin{eqnarray}
    Q_j^0(t)=Q_j(0)\cos(\Omega_j t)+\frac{P_j(0)}{\Omega_j}\,\sin(\Omega_j t).
\end{eqnarray}
Integrating by parts, Eq. (\ref{a_sol1})  can be written as
\begin{eqnarray}
   Q_j(t)=Q_j^0(t)+\frac{c_j}{\Omega_j^2}\left\{
   X(t)-X(0)\,\cos (\Omega_j t)
   -\int_0^t \cos (\Omega_j \tau)\,\dot X(t-\tau)\,d\tau
   \right\}.
   \label{a_sol2}
\end{eqnarray}
Substituting into Eq. (\ref{a_eom1}) yields 
\begin{eqnarray}
\dot P(t)=-\frac{\partial V_{ex}}{\partial X}-\kappa\,X(t)-\int_0^t \Gamma(t-\tau)\, P(\tau)\,d\tau+\zeta(t)-M\,X\, \Gamma(t),
   \label{a_gle1}
\end{eqnarray}
where the dissipation kernel  $\Gamma(t)$ and noise $\zeta(t)$ are
\begin{eqnarray}
    \Gamma(t)=\frac{1}{M}\,\sum_{j=1}^N \left(
    \frac{c_j}{\Omega_j}\right)^2 \cos(\Omega_j t),\quad \zeta(t)=\sum_{j=1}^N c_j Q_j^0(t),
\end{eqnarray}
the effective stiffness $\kappa$ is
\begin{eqnarray}
    \kappa=k_0-\sum_{j=1}^N\left(\frac{c_j}{\Omega_j}\right)^2, 
    \label{a_kappa}
\end{eqnarray}
and $X=X(0)$.
In the limit $N\to\infty$ $\kappa $ is given by expression (\ref{kappa}). Finally, absorbing the term  $-M\,X\, \Gamma(t)$ into noise and using notations of the main text,  we obtain
\begin{eqnarray}
M\ddot X(t)=-\nabla V_*[X(t)]-\int_0^t \Gamma(t-\tau)\, P(\tau)\,d\tau+E_0(t), 
   \label{a_gle2}
\end{eqnarray}
with the PMF 
\begin{eqnarray}
    V^*=V_{ex}+\frac{1}{2}\kappa q^2, 
\end{eqnarray}
and noise 
\begin{eqnarray}
E_0(t)=\zeta(t)-M\,X\,\Gamma(t)=\sum_{j=1}^N c_j\,
   \left[
   \left(Q_j-\frac{c_j}{\Omega_j^2}\,X\right)\,
   \cos(\Omega_j t)+\frac{P_j}{\Omega_j}\,\sin(\Omega_j t)\right],
\end{eqnarray}
where $X=X(0)$, $Q_j=Q_j(0)$, $P_j=P_j(0)$.
Eq. (\ref{a_gle2}) coincides with the Langevin equation (\ref{gle7}).


\end{document}